# Low-noise top-gate graphene transistors

G. Liu[1], W. Stillman[2], S. Rumyantsev[2,3], Q. Shao[1,4], M. Shur[2] and A.A. Balandin[1, *]

[1] Nano-Device Laboratory, Department of Electrical Engineering and Materials Science and Engineering Program, Bourns College of Engineering, University of California – Riverside, Riverside, California 92521 USA

[2] Center for Integrated Electronics and Department of Electrical, Computer and Systems Engineering, Rensselaer Polytechnic Institute, Troy, New York 12180 USA

[3] Ioffe Physico-Technical Institute, Russian Academy of Sciences, St. Petersburg, 194021 Russia

[4] Center for Micro and Nano Technology, Lawrence Livermore National Laboratory, Livermore, California 94550 USA

[*] Corresponding author: email: balandin@ee.ucr.edu, web: http://ndl.ee.ucr.edu

**Abstract**

We report results of experimental investigation of the low-frequency noise in the top-gate graphene transistors. The back-gate graphene devices were modified via addition of the top gate separated by ~20 nm of $HfO_2$ from the single-layer graphene channels. The measurements revealed low flicker noise levels with the normalized noise spectral density close to $1/f$ ($f$ is the frequency) and Hooge parameter $\alpha_H \approx 2\times10^{-3}$. The analysis of noise spectral density dependence on the top and bottom gate biases helped us to elucidate the noise sources in these devices and develop a strategy for the electronic noise reduction. The obtained results are important for all proposed graphene applications in electronics and sensors.

G. Liu, W. Stillman, S. Rumyantsev, Q. Shao, M. Shur and A.A. Balandin, *Low-noise top-gate graphene transistors* (2009)Extraordinary properties of graphene [1-5] such as its extremely high room temperature electron mobility [1-3] and thermal conductivity [4-5] make this material appealing for electronics and sensors. Most of the proposed applications require very low levels of the electronic flicker noise, which dominates the noise spectrum at low frequencies $f < 100$ kHz. The flicker noise spectral density is proportional to $1/f^{\gamma}$, where $\gamma$ is a constant close to 1. The up-conversion of noise, which is unavoidable in electronic systems, results in serious limitations for practical applications. Thus, it is important to investigate the noise level in graphene devices and identify its sources.

Very few studies of the low-frequency noise in graphene devices were reported to date [6-7]. Mostly, the previous works were focused on the back-gated bilayer graphene (BLG) devices. In contract to single-layer graphene (SLG) one can induce a band gap in BLG through the use of an external gate. There has been substantial recent progress in fabrication of graphene transistors with the top gate in addition to the "conventional" back-gate. The back gate is usually separated from the graphene channel by 300 nm of $SiO_2$ required for graphene optical visualization [1-3]. The top gate enables better control of the electronic properties of graphene transistors and may help to achieve the current saturation characteristics [8]. The addition of the top gate frequently leads to the mobility degradation and may increase the noise. However, the double-gate transistor structure allows one for more detail study of the noise sources. In this letter we report the results of the first investigation of the low-frequency noise in the double-gate transistors (also referred to as the top-gate graphene transistor). For our study we selected devices with SLG channels and used $HfO_2$ as the top-gate dielectric.

We prepared SLG flakes with the lateral sizes of ~10 μm by mechanical exfoliation from the bulk highly oriented pyrolytic graphite (HOPG). The number of graphene layers and their quality were verified using the micro-Raman spectroscopy via the deconvolution of the Raman *2D* band and comparison of the intensities of the *G* peak and *2D* band [9-11]. The electron beam lithography (EBL) was used to define the regions for the top gate oxide on the graphene flake and was followed by the low temperature atomic layer deposition (ALD). The thickness of $HfO_2$ directly deposited on top of graphene channel was ~20 nm. A second step of EBL defined the source, drain and the top gate, and was followed by the electron beam evaporation to make Cr/Au (5/60 nm) electrodes. This sequence helped us to avoid possible damage to the contacts during to the long ALD process in the presence of $H_2O$ and precursor environment. Figure 1 shows a schematic of the double-gate graphene transistor structure and an optical microscopy image of a typical device (yellow colour corresponds to the metal contacts; green to $HfO_2$ dielectric and brown to $SiO_2$ dielectric).

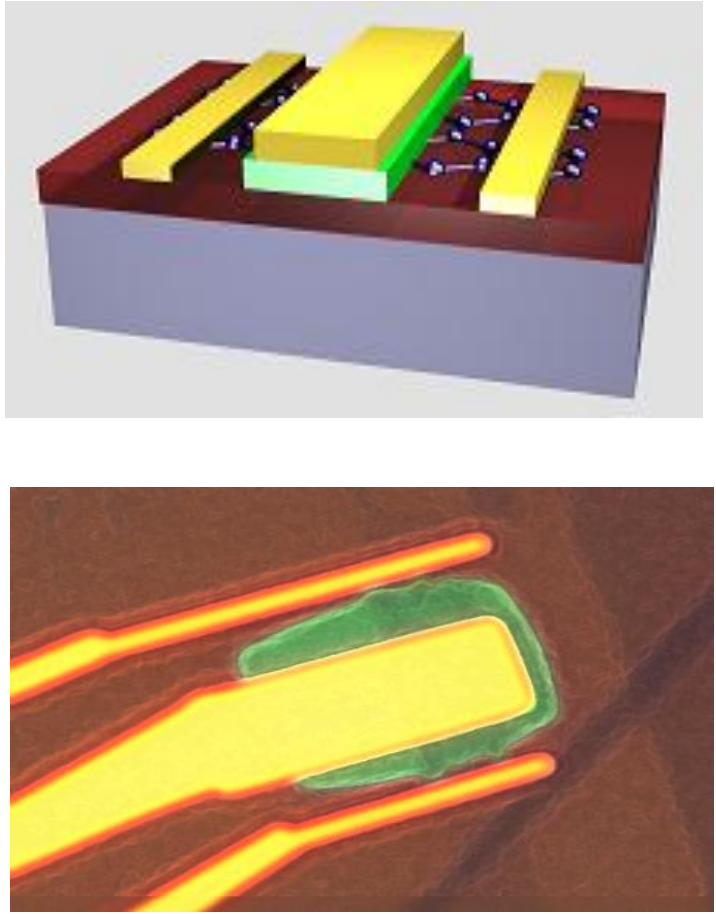

**Figure 1** Schematic of the double-gate graphene transistor (top panel) and an optical image of a typical graphene transistor (brown is $SiO_2$, yellow are metal gates and green is $HfO_2$).

The current – voltage (I-V) characteristics were measured both at UCR and RPI using a semiconductor parameter analyzer (Agilent 4156B). The fabricated devices were robust and retained their I-Vs over the period of testing (about two weeks) at ambient conditions. The top-gate and back-gate functions are shown in Figures 2 (a-b), correspondingly. The Dirac point under the top-gate portion of the graphene transistor channel was $V_D= -1$ V (note the difference with that obtained by tuning the back-gate). The channel conductance was approximately proportional to the gate biases in both cases. The top-gate leakage current in the examined transistors was very small (~1 nA). The mobility for our double-gate transistor with the carriers induced by the back gate was $\mu \approx 1550$ cm$^2$/Vs for electrons and $\mu \approx 2220$ cm$^2$/Vs for holes at room temperature. It was extracted through the Drude formula used previously for graphene devices [1-3]. Following the current-voltage characterization the low-frequency noise was measured with a spectrum analyzer (SRS 760 FFT). The device bias was applied with a "quiet" battery - potentiometer circuit. The drain bias was limited to 50 mV.

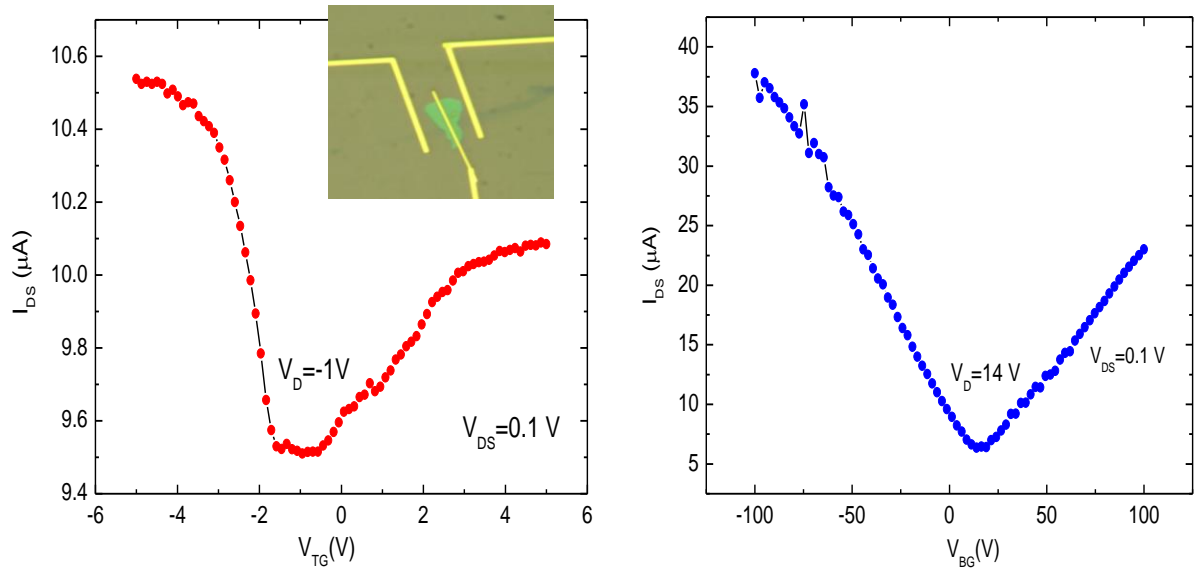

**Figure 2** Drain current as a function of the top gate demonstrating the top-gate action for the back-gate bias $V_{BG}=0$ V (left panel). Drain current as a function of the back gate bias demonstrating the back-gate action at the top-gate bias of $V_{TG} = 0$ V for the same HfO$_2$-graphene-SiO$_2$ double-gate transistor (right panel). The inset shows microscopy image of the measured transistor. The blue colour stripe under the top electrodes is graphene while the green region is HfO$_2$.

The normalized current noise density $S_I/I^2$ for the double-gate graphene transistor is presented in Figure 3. As one can see $S_I/I^2$ is very close to $1/f$ for the frequncy $f$ up to 3 kHz. The $1/f$ noise in electronic devices is often characterized by the empirical Hooge parameter

$$\alpha_H = \frac{S_I}{I^2} fN . \qquad (1)$$

Here $N$ is the number of carriers in the channel estimated as $N=L^2/Rq\mu$ ($L\approx 9$ μm is the source – drain distance, $R$ is the resistance from *I-V* measurements and $q$ is the elemental charge). Using the measured mobility, we obtained the Hooge parameter $\alpha_H \approx 2\times 10^{-3}$. Such values are typical for many metals, semiconductor materials and devices [12]. In this sense the graphene transistors reveal similar noise characteristics as conventional devices.

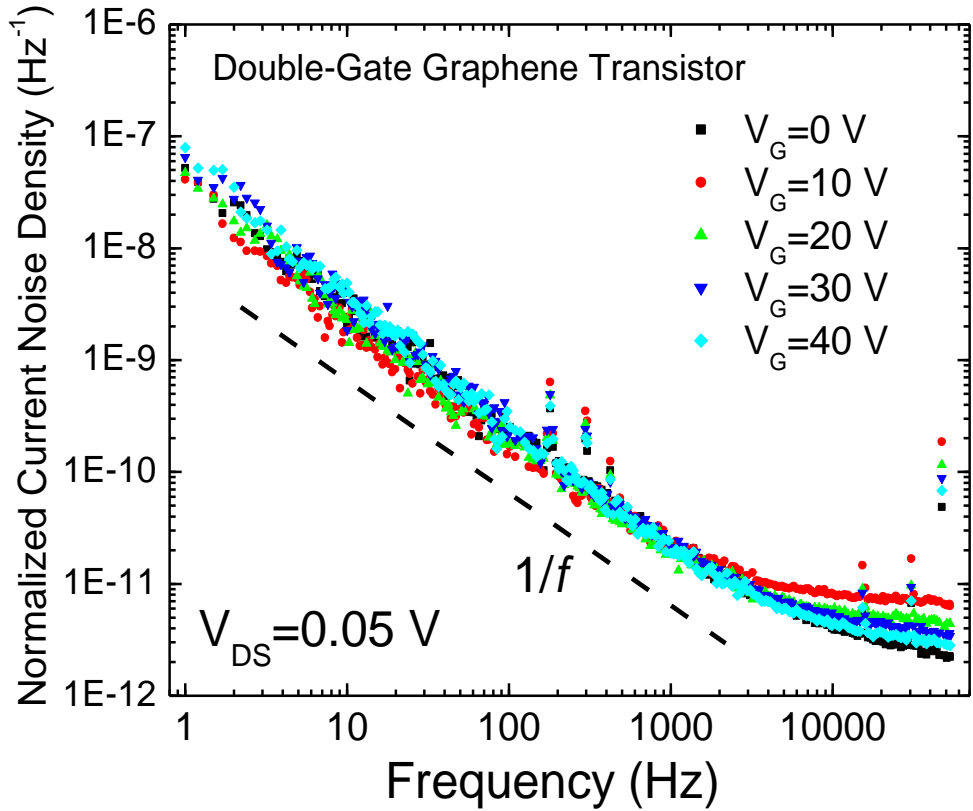

**Figure 3** Normalized current noise spectral density $S_I/I^2$ as a function of frequency *f* for the double-gate graphene transistor. The data is shown for the back-gate bias in the range from zero to 40 V. The *1/f* spectrum is indicated with the dashed line.

The Peransin et al [13] model has been conventionally used for the analysis of the flicker noise sources in transistors [12-14]. Separating the channel resistance into the gated $R_G$ and ungated $R_S$ parts we write for the total channel resistance $R=R_G+R_S$. The noise spectral densities from these two regions are uncorrelated and the measured $S_I/I^2 = S_R/R^2$ can be expressed as

$$S_R = S_{R_G} + S_{R_S} = \frac{\alpha_H R_G^2}{Nf} + S_{R_S} \approx \frac{\alpha_H q \mu R_G^3}{L^2 f} + S_{R_S}. \tag{2}$$

Here Eq. (1) and the expression for the number of carriers were used. This formula can be applied separately for the top and the bottom gates. Since the first term in Eq. (2) depends on the gate bias via $R_G$ and the second term does not depend on the gate bias, there are four possible gate-bias dependencies for $S_R/I^2$. The absence of $S_I/I^2$ dependence on the gate bias suggests that the noise and $R$ are dominated by the contributions from the ungated part of the device channel, i.e. by $R_S$ and $S_{Rs}$. We found that $S_I/I^2$ does not noticeably depend on the top-gate bias $V_{TG}$. For example, at $f=10$ Hz, $V_{DS}=0.05$ V and $V_{TG}$ changing in the examined range (see Figure 2), $S_I/I^2$ stays around $\sim 2\times 10^{-9}$ Hz$^{-1}$. This indicates that the dominant noise contributions in this case do not come from the short-length top-gate region. We did observe a reproducible dependence on the back-gate bias for $V_{BG} \geq 30$ V (see Figure 4).

For small gate biases the $S_I/I^2$ dependence on $V_{BG}$ is weak and cannot be assigned conclusively to one or the other regime distinguished by the Peransin et al [13] model. Although at higher frequencies it seems to tend to $\sim 1/V_{BG}$, which would correspond to the case $R_G>R_S$ and noise is dominated by the gated channel contributions. The strong increase of the normalized noise spectral density with the gate bias (close to $S_I/I^2 \sim (V_{BG})^2$) at high bias may indicate that, while the total resistance is dominated by the gated channel, the major noise contributions come from the ungated parts of the device. We did not observe any clear signatures of the generation – recombination noise (G-R) [12, 15] in the double-gate graphene transistors. In many other nano-scale systems and devices the G-R features became very pronounced [12]. The obtained results contribute to understanding of the low-frequency electronic noise in nano-devices, which often reveal noise characteristics very different from those of conventional devices [16].

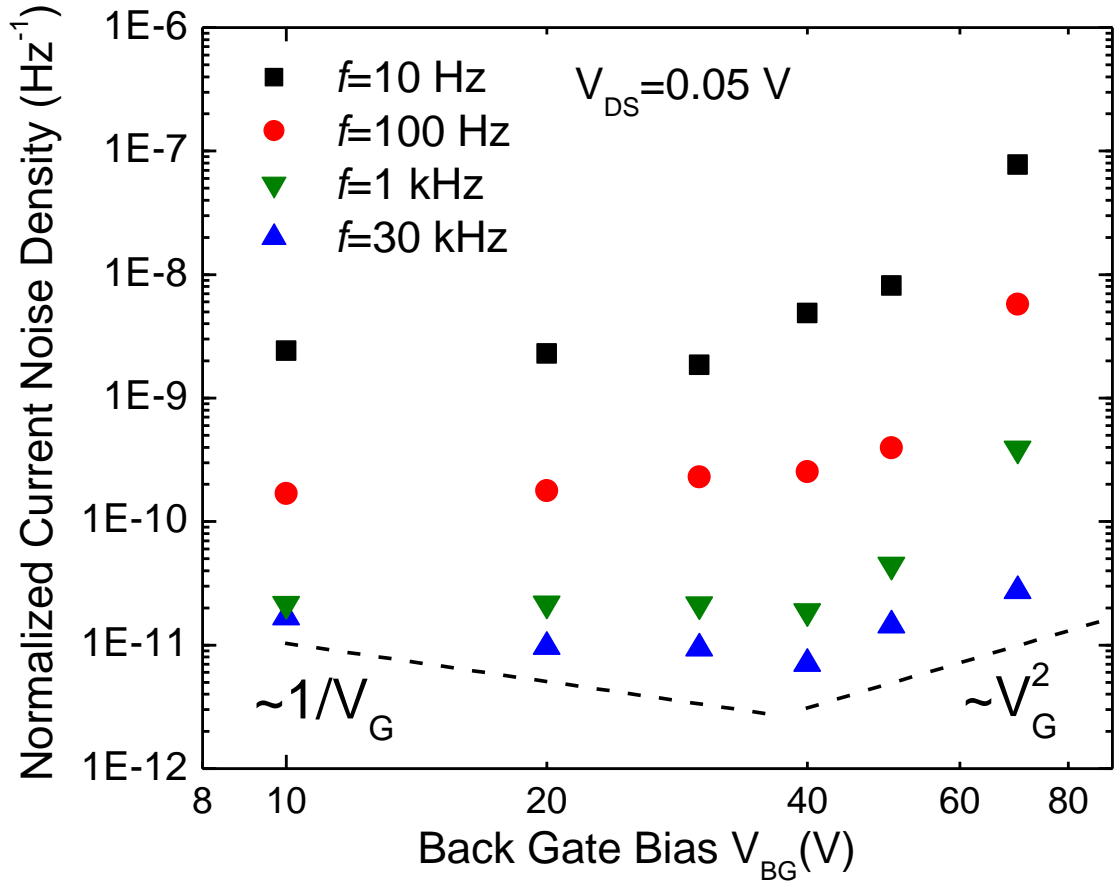

**Figure 4** Normalized current noise spectral density as a function of the back-gate bias for several frequencies. Drain – source bias was kept at 0.05 V.

In conclusion, we have studied experimentally *1/f* noise in the double-gate graphene transistors. The *1/f* noise level in our graphene transistors with the bottom and top gate is rather low with the Hooge parameter $\alpha_H \approx 2 \times 10^{-3}$. The normalized current noise spectrum density $S_I/I^2$ dependence on the bottom and top gates suggests that the contributions from the ungated parts are substantial. Thus, the noise level in graphene transistors can be reduced even further with the improvements in the fabrication technology.

*Acknowledgements*

The work at UCR was supported by DARPA – SRC Focus Center Research Program (FCRP) through its Center on Functional Engineered Nano Architectonics (FENA) and Interconnect Focus Center (IFC) and by AFOSR award A9550-08-1-0100. The work at RPI was supported by the IFC seed funding.